\documentstyle[preprint,prd,aps,psfig,floats]{revtex}
\tightenlines

\def\ombh2{\Omega_{\rm B}h^2}
\def\om0{\Omega_0}

\def\sig8{\sigma_8}

\def\beq{\begin{equation}}
\def\eeq{\end{equation}}

\begin{document}
\pagestyle{empty}

\setcounter{page}{1}
\pagestyle{plain}
\title{Still flat after all these years!}
\author{Elena Pierpaoli,}
\address{Department of Physics and Astronomy, 
University of British Columbia, 6224 Agricultural Road, 
Vancouver, BC V6T 1Z1 Canada\\
and Canadian Institute for Theoretical Astrophysics,Toronto, ON\ M5S 3H8,
Canada}
\author{Douglas Scott}
\address{Department of Physics and Astronomy,
University of British Columbia, 6224 Agricultural Road,
Vancouver, BC V6T 1Z1 Canada}
\author{Martin White}
\address{Harvard-Smithsonian Center for Astrophysics,
Cambridge, MA 02138, U.S.A.}
\maketitle

\begin{abstract}

\noindent
The Universe could be spatially flat, positively curved or negatively curved.
Each option has been popular at various times, partly affected by an
understanding that models tend to evolve away from flatness.
The curvature of the Universe is amenable to measurement, through tests
such as the determination of the angles of sufficiently large triangles.
The angle subtended by the characteristic scale on the Cosmic Microwave sky
provides a direct test, which has now been realised through a combination
of exquisite results from a number of CMB experiments.

\noindent
After a long and detailed investigation, with many false clues,
it seems that the mystery of the curvature of the Universe is now solved.
It's an open and shut case: the Universe is flat!

\end{abstract}
\newpage

Humankind has long pondered whether the world is flat or round -- whether
boats would fall off the Earth and whether one can reach the edge of Creation.
After the coming of General Relativity, the idea of a non-Euclidean
geometry in 3 spatial dimensions began to be applied to the entire Universe.
We imagine space being curved in an imaginary 4th dimension, and that
this curvature can be detected by sufficiently accurate measurements in the
3 dimensions in which we live.

The classic example of a test for non-flatness involves measuring the
angles of a large triangle and determining whether they sum to $180^\circ$,
or to some larger or smaller number.  We know that in the 2-d
analogy, on the surface of the Earth for example,
the angles of spherical triangles always come to more than two right angles.
And for triangles drawn on the inside of a trumpet horn,
an example of a 2-d hyperbolic space, the angles sum to less than
$180^\circ$.  The same test can be applied in 3-d, and, in addition, we
can also consider surface area or volume tests.  The trick is to apply such
tests on a sufficiently large scale, since space is locally extremely flat.
As we shall see, this very test has now been applied using CMB anisotropies,
on a distance scale which is approximately the Hubble radius.

Let us define some terminology.
In an isotropic and homogeneous Universe Einstein's equations imply the
following evolution equation for the scale factor $R(t)$:
\begin{equation}
H^2 \equiv \left( {\dot R \over R} \right) = {{8 \pi G \rho} \over {3}} -
{c^2 \over {\cal R}^2}
\label{eq:exp}
\end{equation}
where   $\rho$ is the total energy density 
and ${\cal R}$ is the radius of curvature of the Universe (we write
${\cal R}\to i{\cal R}$ for open models, and ${\cal R}\to\infty$ for flatness).
Conventionally curvature is measured through the density parameter
$\Omega$, which is defined by
 $\Omega \equiv 8 \pi G \rho / 3 H^2$.  This relates to curvature
because $\Omega - 1 = c^2 / H^2 {\cal R}^2$.
If $\Omega = 1$ then
the Universe is flat, while $\Omega > 1$ and $\Omega <1$ imply a closed and
open geometry, respectively.
There are in principle many ways to measure the total
energy density of the Universe,
but most methods do this indirectly, for example by measuring the density in
different species (obeying different equations of state) separately and then
summing to infer the value of $\Omega$.

{}From the time of Euclid to the time of Einstein, scientists had a firm
understanding of flat geometry, and never doubted that this described the
Universe.
Soon after curved Universes were proposed mathematically,
closed models became the most popular, at least theoretically.
Although there was never good evidence
for believing that $\Omega$ could be significantly bigger than unity,
nevertheless there was some allure for a Universe with finite
volume that also had a very definite future.   $\Omega$ is rather difficult
to measure, and it wasn't until perhaps the 1960s that open Universes gained
observational favour.  Throughout the 1970s and 1980s
there was a split between theoretical and observational cosmologists.
Observers saw only
evidence for relatively low values of $\Omega$, while theorists preferred
to imagine that there might be enough dark matter to
preserve the simplicity of $\Omega=1$.
Theorists continued to hope for $\Omega_{\rm matter}=1$ even when there
was considerable evidence to the contrary.  There was a short period
in the mid 1990s when open universes were popular, and even theorists got
excited by inflationary open models.
Now it appears that after many false trails we are once again
back on the road to flatness.

Flatness does of course imply extreme fine tuning in the early Universe
\cite{Dicke}, since $\Omega=1$ is unstable.  There has been much philosophical
debate over whether it is a serious problem that the Universe can have
contrived to avoid diverging much from flatness.  On the one hand there
are those who contend that since $\Omega={\rm constant}$ has only 3 solutions,
0, 1 and $\infty$, then faced with that choice there is only one reasonable
option.  Others argue that this only makes sense if one invokes a mechanism
for fine tuning the Universe to $\Omega=1\pm\epsilon$ at early times, and that
inflation comes to the rescue.  Yet others would argue that the anthropic
principle saves us from living in Universes where $\Omega$ departs too far
from unity.  We prefer to take the pragmatic approach -- let us wait to see
what the curvature of the Universe turns out to be, and {\it then\/} worry
about the explanation.  It appears that the wait is now over.  $\Omega$ is
indeed very close to unity.  So now we can worry about what it all means!

The newly applied test, using Cosmic Microwave Background anisotropies, has
the appeal of being a much more direct
measurement of $\Omega$, regardless of the specific kinds of
energy or matter which contribute to the $\Omega$ census.
The only loop-hole is that one must first be confident that the test is
being applied within the correct cosmological model.
Once that has been established, then
$\Omega$ can be constrained by looking at the position
of the peak in the CMB power spectrum, which is expected to exist
in most popular cosmological models.
CMB anisotropies are measured by looking at the variation of mean square
temperature fluctuation with angular scale -- conventionally one uses
the power spectrum versus multipole $\ell$, where $\ell\sim 1 / \theta$.
Some examples of power spectra are shown in Fig.~1.

\begin{figure}
\centerline{\psfig{figure=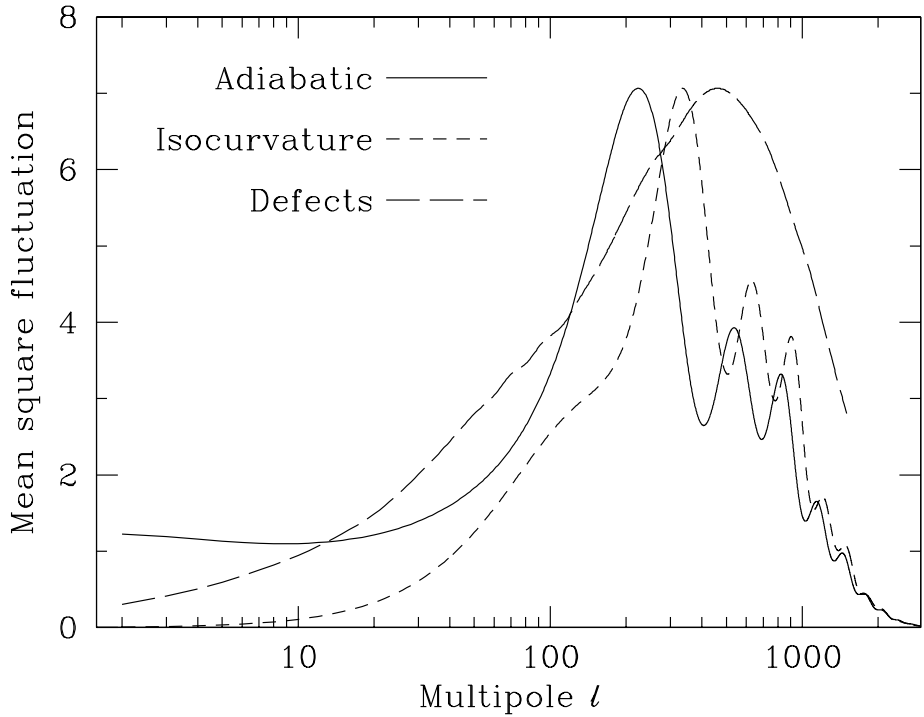,height=11cm}}
\caption{Theoretical predictions for 3 classes of model.  All of these have
a cosmological constant, and flat geometry, and they have been normalized
to have the same peak height, regardless of their large angle normalization.
The isocurvature model is CDM-isocurvature.  Both adiabatic and isocurvature
models correspond to scale-invariant initial conditions.  The defect model is
a recent cosmic string calculation [6].
}
\end{figure}

The CMB triangle test involves using a standard ruler on the last scattering
surface and then attempting to measure its angular size.
The ruler here is the sound horizon at last scattering, corresponding to
the size of the fundamental acoustic mode at that epoch.
The lengths of the two longest sides of the triangle are the
distance to the last scattering surface, which is essentially fixed.
With all 3 sides of the triangle having their lengths established,
all that remains is
to measure the angle subtended by the shortest side, the standard ruler
on the last scattering surface, and we have determined whether or not the
Universe is flat \cite{test}.

In the most popular models, $\ell_{\rm peak}\simeq220$ in a flat Universe.
If $\Omega$ is changed, the main effect is that the location of the 
peak shifts according to $\ell_{peak} \propto \Omega^{-1/2}$.
Of course the height of the peak is also affected,
but in a manner which strongly depends on
other parameters, while the position of this main acoustic peak is a robust
measure of $\Omega$ alone.

In order to apply the CMB flatness test, it is important first to establish
the framework in which to perform the test.  There are 3 separate classes
of model for generating cosmological perturbations:
\begin{itemize}
\item adiabatic initial conditions;
\item isocurvature initial conditions;
\item topological defect sources.
\end{itemize}
As shown in Fig.~1, all of these ideas for the origin of perturbations
can produce at least one peak in the power spectrum. 
To understand the basic differences, imagine a scenario in which 
fluctuations are laid down at very early times in
a way which is apparently acausal.  In other words, modes over all relevant
length scales begin with the same phase, irrespective of whether there
has apparently been enough time for the establishment of causal contact.
Inflation carries this off by having the Universe effectively
expand faster than the speed of light \cite{Inflation}
-- but any initial condition
synchronized over a sufficiently large volume would be equivalent.
Such perturbations can then
be either of the adiabatic or isocurvature type or, in principle, a
combination of the two.  `Adiabatic' means that the entropy per
species is unperturbed, while `isocurvature' means that the total energy
density is unperturbed.
CMB anisotropies can be thought of in terms of a driven harmonic
oscillator \cite{HuSS} with the driving force being the gravitational
perturbation.  For adiabatic modes, we have a driving force which
already exists when the modes start to oscillate, shortly after they come
inside the horizon, whereas for isocurvature modes the driving force is
growing from zero.  The phases of the oscillating fluid are captured in a
snapshot at the last scattering epoch and hence we find
that the power spectra of adiabatic and isocurvature modes are in
anti-phase with each other (Fig.~1)\cite{HuSW}.

Topological defect models generate density perturbations as the field re-orders
itself on roughly the horizon scale.  These modes are therefore causal.
We can think of a model like cosmic strings as arising from a number of
patches on the sky, each of which is like an isocurvature model which chose
some random number for the initial phase of the modes.  The final power
spectrum is the incoherent average of a number of shifted isocurvature
power spectra, and thus shows a single broad peak (Fig.~1).

In Fig.~1 we show an example of a power spectrum from each of the three main
families of models that have been considered for generating structure.
We have specifically chosen models dominated by a cosmological
constant (to be consistent with currently popular ideas).  The curves have
been scaled to have the same peak height.
\nocite{PogVac}
These three generic predictions look quite different.
Note that it is possible to argue for other sources of anisotropy at the
smallest $\ell$s, and one should focus one's attention at the higher
multipoles.

\begin{figure}
\centerline{\psfig{figure=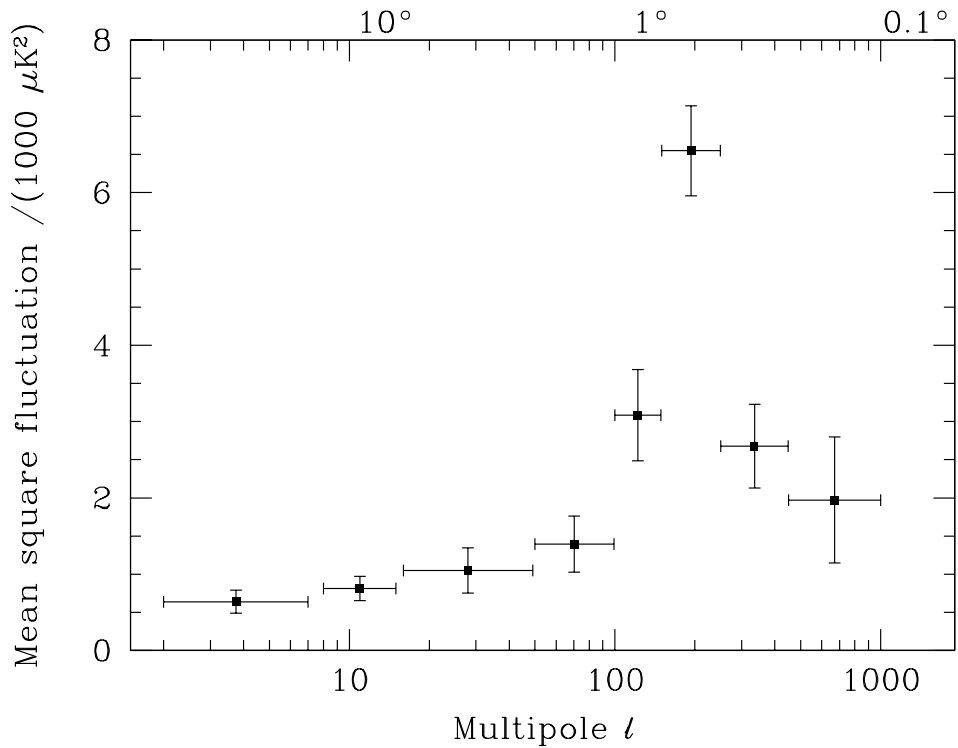,height=11cm}}
\caption{Binned CMB anisotropy power spectrum obtained from a combination of
current experimental data [9].}
\end{figure}

Evidence has been accumulating from CMB experiments, as well as the
relative normalization with the matter fluctuation power spectrum,
which indicate that the fluctuations are adiabatic.
Some individual CMB data sets give evidence for a localised peak
\cite{Meletal}.  However, it is the compilation of all available data
that gives a convincing demonstration\cite{peak}.
To this end,
we have calculated a binned anisotropy power spectrum, which we plot in Fig.~2 
\cite{Science}.
The resulting power spectrum is flat at large angles, with a gradual rise to 
a prominent peak a little below $1^\circ$, with a clear decrease thereafter. 
This is precisely the shape predicted by inflationary-inspired adiabatic
models. 

Although we can imagine that small components of isocurvature or
defect-produced perturbations may still be possible \cite{Elena},
it is apparent is that overall the adiabatic models are in very good shape.
Therefore, we can assume the adiabatic family, and use the position of the
peak in $\ell$ to constrain the geometry of the Universe within the
framework of adiabatic models.
We do this rather simply by taking a cosmological constant dominated model,
of the sort which fits a wide range of present cosmological data, and we
rescale the $\ell$ scale by $\Omega^{-1/2}$ to find a range
of acceptable values of $\Omega$.  This provides a 95\% confidence region
$0.79<\Omega<1.17$, or approximately $\Omega=1.0\pm0.1$.
It thus appears to be settled that the Universe is
very close to flat -- much closer than would be implied by the amount of
visible or dark matter inferred from studies of galaxies, typically
$\Omega_{\rm matter}\simeq0.3$.

The next wave of CMB experiments should determine more definitively whether
the perturbations are adiabatic, by measuring the power spectrum at angular
scales where oscillations are predicted.  The position of the first peak should
soon be determined more precisely.  In addition, other tests, including
those from distant supernovae, should help determine the constituents
that make up $\Omega$.

It may be that one cosmic mystery has been solved, and it is tempting to
regard this as something akin to a proof that the Universe once went through
a period of inflation.  However, new mysteries have been uncovered: what
exactly was the physics which led to these adiabatic perturbations; what
is this `dark energy' which makes the Universe flat? and why did the Universe
contrive to contain baryons, dark matter and dark energy in similar
proportions?

\bigskip

\acknowledgements
This research was supported by the Natural Sciences and Engineering 
Research Council of Canada, and by the US National Science Foundation.

\end{document}